\documentclass[twocolumn,showpacs,preprintnumbers,amsmath,amssymb]{revtex4}
\usepackage{graphicx,epsf}
\begin{document}

\title{Can randomness alone tune the fractal dimension?}
\author{M. Kamrul Hassan$^{1,2}$ and J\"urgen Kurths$^1$}
\affiliation{
$^1$University of Potsdam, Department of Physics, Postfach 601553, D-14415 Potsdam, Germany\\
$^2$ University of Dhaka, Department of Physics, Theoretical Physics Division, Dhaka 1000, Bangladesh
}
\email{khassan@agnld.uni-potsdam.de}

\begin{abstract}%
We present a generalized stochastic Cantor set by means of a simple {\it cut and delete process} and discuss the
self-similar properties of the arising geometric structure. To increase the
flexibility of the model, two free parameters, $m$ and $b$, are introduced which tune the 
relative strength of the two processes and the degree of randomness respectively. In doing so, we have identified 
a new set with a wide spectrum of subsets produced by tuning either $m$ or $b$. Measuring the size of the  
resulting set in terms of fractal dimension, we show that the fractal dimension increases with increasing order
and reaches its maximum value when the randomness is completely ceased.  

\end{abstract}


 \maketitle


\section{Introduction}

The history of describing natural objects by geometry is as old as science itself. Traditionally,
this has involved Euclidean geometry that restricts the natural boundary of our imagination
to lines, circles, cones, spheres, cuboids and so on. However nature is not restricted to
Euclidean shapes only. Instead, most of the natural objects we see around us are so irregular
and complex in shape that they can be described as geometrically chaotic, since they are not just merely complex
but often contain different degrees of complexity. In 1975,
Mandelbrot introduced the concept of fractal geometry to characterize these geometric monsters quantitatively.
This extended our horizon of understanding and enables to appreciate that there 
exists some kind of order even in these seemingly complex and disordered geometric structural patterns. 
Often fractals are perceived as some brightly-colored computer generated patterns. 
This is due to the lack of enough analytically solvable theoretical models. The coastline of an island,
 a river network, the structure of a cabbage or broccoli, or even the networks of nerves and blood
vessels in the normal human retina can be best described as fractals. Yet, more than twenty five years
after they were first introduced, there is no generally-accepted definition of a fractal. 
However, it is loosely defined as a shape made of parts similar to the whole, in some sense. 
It is typically quantified  by a non-integer exponent called the fractal dimension that can uniquely
characterize the geometric structure. This definition immediately confirms the existence of scale invariance, that is,
the objects look the same on different scales of observation.
To understand fractals, their physical origin and
how they appear in nature we need to be able to model them theoretically.
This is the main motivation of our present work.

The simplest way to construct an ideal fractal is by repeating a given operation over and over 
again deterministically. The classical Cantor set is such a simple text book example
which is typically used for learning about the self-similar properties of an ideal fractal. 
It is created by dividing a line into $n$ equal pieces, removing $(n-m)$ of the parts created
and repeating the process with the $m$ remaining pieces {\it ad infinitum} \cite{kn.mandelbrot}. 
However, it is well understood
that in our world nothing is stationary or strictly deterministic and all natural complex objects, 
even intelligence and life, occur through
some kind of evolution where randomness is an essential ingredient. 
That is, nature favors randomness and fractals in nature appear through a continuous kinetic process, whereas
the classical Cantor set is discrete in time. The simplest way to incorporate these essential
ingredients is by generalizing the Cantor set. For example, instead of dividing an interval into $n$ equal
pieces, one can divide it randomly into $n$ pieces and throw away $(n-m)$ of these also randomly \cite{kn.hassan1}.

The notion of random fractal has been widely used in almost every disciplines of science both theoretically and 
experimentally since its inception. Yet, the mechanism by which nature creates fractals and the relationship 
between the degree of order and the fractal dimension is poorly understood. 
In this article, we introduce a stochastic process that can generate a continuous spectrum of fractals
(ranging from random to nonrandom) controlled by some
intrinsic tuning parameters. This may be considered as a natural kinetic counterpart of the classical
Cantor construction and which can be a potential candidate in order to
understand essential governing rules of creating complex geometric objects. 
We present a generalized version of the classical Cantor set. The rules of the process
can be described as follows. The process starts with an initiator of the unit interval 
[0,1] and a generator which divides this interval into two pieces randomly, deleting
some parts from either sides of both pieces at each time step. We introduce
two intrinsic parameters to determine the degree of randomness and the rate at which a given 
operation is to be repeated in order to create a fractal.

The construction of this generalized Cantor set is not at all
pedagogical. One immediate and potential application of the present model
is the kinetics of the irreversible and sequential breakup of particles 
that occurs in a variety of physical processes and which has important applications in science and technology.
 These include erosion \cite{kn.has0}, grinding and crushing of solids\cite{kn.has1}, polymer 
degradation and fiber length reduction \cite{kn.has2}, breakup of liquid droplets 
\cite{kn.has3}, etc. to name just a few. In recent years, there has been an increasing 
interest in studying fragmentation, allowing variations that increase the flexibility of 
the theory in matching conditions of real phenomena such as the extension to higher
dimension \cite{kn.has4}, agglomerate erosion \cite{kn.has0}, mass loss \cite{kn.has5}, 
volume change \cite{kn.has6} and fragmentation-annihilation \cite{kn.has7}. 
The generalized Cantor construction, as we formulated it, can 
describe the kinetics of fragmentation with a continuous mass loss
\cite{kn.hassan2,kn.has5}. This is relevant in the kinetics of fragmentation of particles 
where mass loss might occur due to evaporation, oxidation, 
sublimation, dissolution, melting etc., or even in the Yule-Furry process of cosmic shower theory 
where energy loss occurs due to collision \cite{kn.cos}.

Another potential application of the present work is the first order phase transformation via the kinetics
of nucleation and growth processes. The random deposition of cuts in the 
generalized Cantor set can be thought of as the random nucleation of size-less seeds on the substrate (metastable phase).
On the other hand, the continuous removal of parts is in fact equivalent to the decrease of substrate 
due to the growth of seeds (stable phase).
The nucleation of a stable phase usually occurs due to large thermal
fluctuations in the metastable phase and the subsequent growth occurs due to the competition between the 
surface tension and the free energy density difference between the stable and metastable phase. The classical 
work of such processes has been done independently by Kolmogorov, Johnson, Mehl and Avrami  (KJMA model)
\cite{kn.kolmogorov} and has been extensively used to describe non-equilibrium phenomena during 
phase transformation \cite{kn.krapivsky}. 
In this model the nucleation of a stable phase,
within the metastable phase, occurs at random obeying the Poisson stochastic process in space. In other words
nucleation centers are homogeneously distributed throughout the entire metastable phase. Once nucleated,
the stable phase keeps growing isotropically until it comes into contact with another similarly growing stable phase 
thus forming an interface.


Interesting questions arise from the present work: (i) What is the
role of the fractal dimension during the pattern formation? (ii) Is there
any relation between the fractal dimension and the degree of order? (iii) What
are the relevant parameters to tune the degree of order and what are
their physical meaning? The present work is an attempt to answer these
questions. 
 
\section{Formulation of the cut and delete process} 

In order to create a complex geometric objects here we give the general scheme for the formulation of the
{\it cut and delete} model. To do so, we define the interval size distribution function or the concentration
$c(x,t)$ so that $c(x,t)dx$ describes the number of intervals of size $x$ at time $t$ in the interval 
range $x$ and $x+dx$. The distribution function $c(x,t)$ then must have the following form of evolution equation
\begin{equation}
{{\partial c(x,t)}\over{\partial t}}  =  \left. {{\partial c(x,t)}\over{\partial t}}\right |_{{\rm cut}}+
\left. {{\partial c(x,t)}\over{\partial t}}\right |_{{\rm delete}}.
\end{equation}
The first term on the right hand side represents the irreversible and sequential cut process which can be described
by the following master equation
\begin{eqnarray}
\left. {{\partial c(x,t)}\over{\partial t}}\right |_{{\rm cut}} & = &  -
c(x,t)\int_0^xdy F(y,x-y) \nonumber \\ & +  &  2\int_x^\infty dy c(y,t)F(x,y-x). 
\end{eqnarray}
Here $F(x,y)$ is the breakup kernel describing the rules and the rate ($a(x)=\int_0^xdy F(y,x-y)$) 
at which an interval of size $x+y$ breaks into sizes $x$ and $y$. The two terms in Eq. (2) describe the 
destruction and creation of interval of size $x$ due to the deposition of cuts on size $x$ and larger 
than $x$ respectively. The second term of Eq. (1) on the other hand, represents the deletion of parts of an 
interval from either ends soon after a cut is 
deposited. The deletion of parts continues until these parts encounter another cut or the 
interval becomes a dust like interval;
thereby stopping the loss of their masses. This deletion process can be expressed in the following differential
form 
\begin{equation}
 \left. {{\partial c(x,t)}\over{\partial t}}\right |_{{\rm delete}}=
{{\partial\big(q(x)c(x,t)\big) }\over{\partial x}}. 
\end{equation} 
where $q(x)$ describes the rate at which an interval of size $x$ is lost by the deletion process.   

\subsection{Model I: Random scission model}

We first consider here the simplest possible model of this kind, namely 
\begin{equation}
F(x,y)=1.
\end{equation}
This is typically known as the {\it random scission} model \cite{kn.ziff} as it assumes that 
 every points of the interval are equally likely to break. 
That is, the present model describes the following {\it recursive cut and delete process}.
We start with an interval of size $x+y$.
At each time step, we choose a cut point with a uniform probability density and produce two smaller intervals
of size $x$ and $y$. As soon as the new intervals are born, they start shrinking from either side, although the
mechanism whereby they shrink would depend on the choice of $q(x)$. 
In general, q(x) is assumed to follow a 
power law form with respect to particle size i.e. $q(x)\propto x^\gamma$.
However, once the choice for the breakup kernel is made, the exponent $\gamma$ is
strictly restrictive, due to dimensional consistency. Therefore, in the present context 
it is essential to choose $q(x) =mx^2$ where $m$ is a positive, real and dimensionless constant.
The rate equation for the distribution function $c(x,t)$ then becomes
\begin{equation}
{{\partial c(x,t)}\over{\partial t}}=-xc(x,t)+2\int_x^\infty c(y,t)dy+m{{\partial x^2c(x,t)}\over{\partial x}}.
\end{equation}
Instead of trying to solve the rate equation directly, we attempt to solve the rate equation for $M_n(t)$, 
the $n^{{\rm th}}$ moment of $c(x,t)$, defined as $M_n(t)=\int_0^\infty x^nc(x,t)dx$ with $n>0$.
Incorporating this definition into Eq. (5), we obtain
\begin{equation}
{{dM_n(t)}\over{dt}}=-\left [{{n^2+n(1+1/m)-1/m}\over{(n+1)}}\right ]M_{n+1}(t).
\end{equation}
Note that for $m=0$ the total mass or the first moment $M_1(t)$ is a conserved quantity. However, for 
$m>0$ the system violates the mass conservation due to the continuous mass removal by the deletion process.
The interesting feature of the above equation is that for $m > 0$, there are infinitely many conserved
quantities depending on $m$. We can find these conserved quantities by simply inspecting the value of $n=d_f$ for 
which the moment $M_{d_f}(t)$ is independent of time which is obtained by setting $dM_n(t)/dt=0$. 
The problem then rests on solving a quadratic equation in $n$. The
real positive root of the equation then is
\begin{equation}
d_f=-{{1}\over{2}}\big(1+{{1}\over{m}}\big)+{{1}\over{2}}\sqrt{\big(1+{{1}\over{m}}\big)^2+{{4}\over{m}}}
\end{equation}   
while the other root is $\alpha=-(d_f+1+{{1}\over{m}})$ thus $\alpha<0$. We can re-write the rate equation for 
$M_n(t)$ as
\begin{equation}
{{dM_n(t)}\over{dt}}=-{{(n-d_{f})(n-\alpha)}\over{(n+1)}}M_{n+1}(t).
\end{equation}
Iterating the above equation enables us to obtain
all the derivatives of $M_n(t)$ at $t=0$. Using this and the monodisperse initial condition 
into the Taylor series expansion of $M_n(t)$ about $t=0$ gives the explicit and exact solution for $M_n(t)$
\begin{equation}
M_n(t)=~_2F_1(n-d_f,n-\alpha;n+1;-t)
\end{equation}
where $~_2F_1(a,b;c;d)$ is the hypergeometric function. In principle, finding a proper integral transformation
of Eq. (9) and comparing it with the definition of $M_n(t)$ would enable us to recover the exact solution for the 
distribution function $c(x,t)$ (widely known as Charlesby's method \cite{kn.charlesby}). However, in the present work we are more interested in $M_n(t)$ and its asymptotic
temporal behavior than in $c(x,t)$. The asymptotic expansion of the hypergeometric
function \cite{kn.luke} reveals that the exact time dependence of $M_n(t)$ is $M_n(t)\sim t^{-(n-d_f)}$.
The asymptotic decay of $M_n(t)$ is thus linear in $n$ which implies that the system obeys a simple scaling. This will 
be shown vigorously in the following section. 

\subsubsection{Scaling theory}

A closer inspection of Eq. (5) immediately reveals that there are two governing parameters
$x$ and $t$ that completely describe the governed parameter $c(x,t)$. This is due to the fact that the
dimension of the two variables $x$ and $t$ are interlocked by the choice of the breakup kernel and through
the rate equation itself. That is, $x$, the co-factor of $c(x,t)$ of the first term on the right
hand side of Eq. (5), must bear the dimension inverse of time $t$. This enables us to choose any of the two as 
an independent parameter while the other can be expressed in terms of this. That is, we can define $\xi=tx$ as 
a dimensionless quantity. So, if we assume $t$ to be the independent quantity then $c(x,t)$ can be expressed
in terms of $t$ alone i.e.
\begin{equation}
c(x,t)\sim t^\theta \phi(\xi).   
\end{equation}
Here $\theta$ is yet an undetermined exponent and must take the value for which $t^\theta$ bears the dimension of
concentration or the distribution function $c$. This means that all plots of ${{c}\over{t^\theta}}$ vs $xt$
for any initial condition must fall onto each other which is the hallmark for the existence of scaling typically
known as {\it data collapse formalism} \cite{kn.stanley}. 
Substituting the scaling ansatz (Eq. (10)) into the definition of $M_n(t)$ and demanding
that $M_{d_f}$ be a conserved quantity, one immediately obtains $\theta=1+d_f$.   

\subsubsection {Fractal analysis}



\begin{figure}[!htb]
\centerline{\includegraphics[width=8.0cm]%
{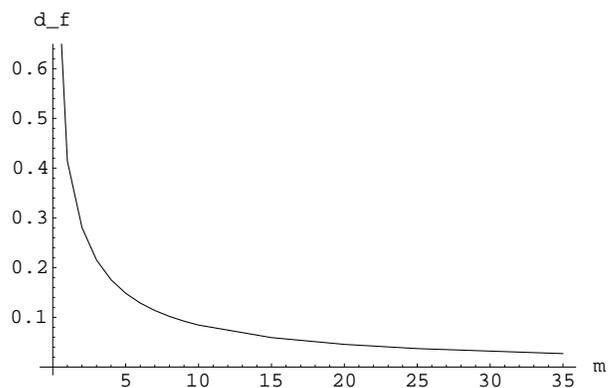}}
 \caption{Fractal dimension versus mass removal strength $m$.}
  \label{fig:1}
\end{figure}

Associating broken objects with disorder is perhaps the most immediate and natural thought that cross 
our mind. Indeed, an assembly of fragmented intervals can be considered as one of the simplest 
yet most common example of a disordered system. The existence 
of scaling is already a proof that an assembly of such sequentially broken intervals has some degree of regularity.   
Owing to the random nature of the present model and due to the mass removal term, it is clear 
that when the process evolves {\it ad infinitum}, it will not create the whole set that 
describes a line. Rather, it will produce a subset of the full set (line). 
In order to measure the size of the set created in the long time limit,
we define the average interval size as
\begin{equation}
\delta(t)={{\int_0^\infty x c(x,t)dx}\over{\int_0^\infty c(x,t)dx}}\sim t^{-1}.
\end{equation}
This shows that the mean or typical interval size decreases in time algebraically. 
We can use this as a yard stick and count the number of the segments we need to cover the resulting set. 
We find that the number of  yard stick $N(\delta)$ scales as  
\begin{equation}
\lim_{\delta\longrightarrow 0}N(\delta)\sim \delta^{-d_f}
\end{equation}
where $N=\int_0^\infty c(x,t)dx$. The exponent $d_f$ is known as the fractal dimension 
or the Hausdorff-Besicovitch dimension of the resulting set created by the {\it cut and delete model}. The distance 
between the points in space is the key to the definition of the  Hausdorff-Besicovitch dimension \cite{kn.feder}.
Note that $d_f$ is the root of the quadratic equation in $n$ obtained by setting
${{dM_n(t)}\over{dt}}=0$ (a condition for stationary solution). The expression for $d_f$, Eq. (7), 
reveals that as the value of 
$m$ increases, the fractal dimension decreases very sharply and in the limit $m \longrightarrow \infty$, 
$d_f \longrightarrow 0$. Fig. (1) does indeed show that as $m$  increases the 
strength of the mass deletion term increases
with respect to the terms responsible for describing the cut processes. This means that as $m$ increases, the size of the
arising set decreases sharply due to the fast disappearance of its member. While as $m\longrightarrow 0$, 
$d_f \longrightarrow 1$, that is, we recover the full set from which the subset is derived.

\subsection{Model II: Gaussian cut and delete model}

Next we treat the following quadratic type of breakup kernel
\begin{equation}
F(x,y)=(xy)^{b-1}.
\end{equation}
This is a generalized version of the previous model and it describes the 
following {\it recursive cut and delete process}. We start with an 
interval $(x+y)$ and at each time event we choose a cut point with a Gaussian probability density if $b>1$. 
The spread of the density of the cut around the center of the interval depends on the values of $b$ as we will
show later. Of course we recover the previous random scission model when $b=1$.
Our principal goal of studying this model is to understand the role of $b$ during
the {\it cut and delete process}. In order to extract the role of $b$, it is  essential
to check whether the relative strength between the {\it cut and delete process} changes as we change
the $b$ value. Note that the breakup rate $a(x)$ for this model is $a(x)=px^{2b-1}$
where $p=[\Gamma(b)]^2/\Gamma(2b)$. It is thus clear that as $b$ increases so does
the relative strength of the cut process with respect to the delete process if we do not 
tune the $m$ value accordingly. That is, to judge the role of $b$ we ought to give  
equal weight to all the terms of Eq. (1) so that each of them can compete on
an equal footing. This can only be done if we set $q(x)=px^{2b}$ (i.e. $m=p$) and hence the relative 
strength between the two terms of
Eq. (1) stays the same as the value of $b$ increases. The rate equation for $M_n(t)$ then is
\begin{equation}
{{dM_n}\over {dt}}= -\Gamma(b)\left({{[\Gamma(b)]}\over{\Gamma(2b)}}(n+1)-{{2
\Gamma(n+b)}\over{\Gamma(n+2 b)}}\right )M_{n+2b-1}.
\end {equation}
The problem now rests on finding the $n$ value 
for which the moment becomes a time independent quantity.
From the previous fractal analysis we learnt that the real positive root of the polynomial equation 
(obtained by setting $dM_n(t)/dt=0$) is in fact 
the fractal dimension of the resulting set. Applying the scaling theory based on dimensional analysis, like
we did in the previous section, gives $\theta=(1+d_f)z$ where the kinetic exponent is $z=1/(2b-1)$. 
A detailed numerical survey reveals that the fractal dimension increases monotonically with increasing 
$b$ (see Fig. (2)). In order to find the fractal dimension in the limit $b\longrightarrow \infty$,
we can use the Stirlings approximation in the resulting polynomial equation. 
In doing so we obtain the following equation 
\begin{equation}
\ln[n+1]+(n-1)\ln[2]=0
\end{equation} 
and find $d_f=0.45699956$ is the only real positive root and hence is the fractal dimension $d_f$.
In an attempt to search for the role of $b$ we study the following model
\begin{equation}
F(x,y)=(x+y)^\gamma \delta(x-y).
\end{equation}
It describes that at each time an interval is chosen and the cut is placed exactly in the middle to produce
two equal sized smaller intervals. It is note worthy to mention that the homogeneity index gamma
only tunes the rate $a(x)=x^\gamma/2$ and does not play any role in controlling the location of each cut on the interval
{\it vis-a-vis} the fractal dimension. The exact value of $\gamma$ is not important provided $\gamma>0$ and therefore
we can choose $\gamma=2b-1$ without loss of generality. Once again the dimensional consistency requires 
that $q(x)=x^{2b}/2$ while the factor $1/2$ is to account for the equal weight in the {\it cut and delete process}. 
Introducing this model into Eq. (1) and inserting the definition of moment we obtain the following 
rate equation for $M_n(t)$
\begin{equation}
{{dM_n(t)}\over{dt}}=-\Big({{(n+1)}\over{2}}-{{1}\over{2^n}}\Big)M_{n+2b-1}(t).
\end{equation}


\begin{figure}[!htb]
\centerline{\includegraphics[width=8.0cm]%
{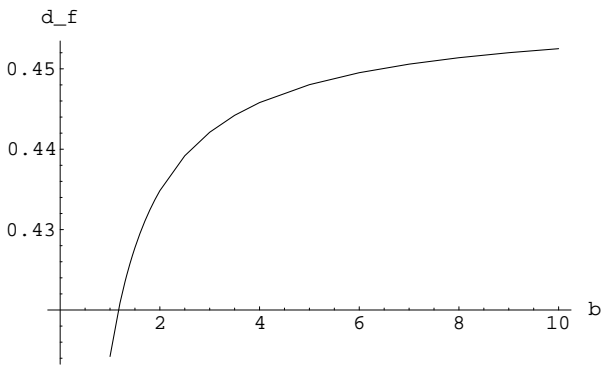}}
 \caption{Fractal dimension versus $b$.}
\label{fig:2}
\end{figure}

In order to find the fractal dimension for this process, we set 
the numerical factor on the right-hand side of this equation equal to 
zero and then take the natural logarithm on both sides to obtain the $n$ value for which $M_n(t)$ 
is time independent. In doing so, we arrive at the same functional equation for $n$ as 
we found for the Gaussian model with $b \longrightarrow \infty$, namely Eq. (15). 

This is in fact the main message of the present work.
It shows that the kernel $F(x,y)=(xy)^{b-1}$ behaves exactly in the same fashion as for 
$F(x,y)=(x+y)^{2b-1}\delta(x-y)$. We thus find that in the limit $b \longrightarrow 
\infty$, the resulting distribution of points is a set with fractal dimension 
$D_f=0.45699956$, which is a strictly self-similar fractal as randomness is ceased by
dividing the intervals into equal pieces.  We are now in a position to give a physical 
picture of the role played by $b$. First of all, the model with $b=1$ 
describes that the frequency curve of placing cuts about the size of the fragmenting 
particles is Poissonian in nature. Consequently, the system enjoys the maximum randomness 
and the corresponding fractal dimension is $d_f=0.414213$. For $b >1$, the 
frequency curve of placing cuts about the size of the fragmenting particles is Gaussian 
in nature, meaning that as the value of $b$ increases, the particles are increasingly more likely to 
break in the middle than on either end. That is, as $b$ increases, the variance 
decreases in such a manner that in the limit $b \longrightarrow \infty$, the variance 
of the frequency curve becomes infinitely narrow, meaning a $\delta$-function distribution 
for which the fragments are broken into two equal pieces. Therefore, there is a continuous spectrum of fractal 
dimensions between $b \rightarrow 0$ when $d_f=0.414213$ and $b \longrightarrow 
\infty$ when $d_f=0.45699956$ (see Fig. (2)).

\vspace{0.5cm}

\section{Discussion}

In this work we have presented a {\it cut and delete process} that can capture the notion of the Cantor set construction.
It has potential applications both in the kinetics of fragmentation and in the kinetics of nucleation and growth
 phenomena. We have found that in the long time limit the system
creates a geometric pattern which is statistically self-similar. To quantify the resulting geometric 
structure, we invoke the idea of fractal geometry and show that the fractal dimension is strictly
 dependent on the strength of the mass removal term $m$. Thus by tuning this $m$, we obtain a 
fractal dimension of any value between $0<d_f\leq 1$. We have further generalized our model allowing a parameter
$b$ that can tune the degree of randomness by controlling the way the cuts are deposited. 
This is done in an attempt to find a precise answer of how fractal
dimension changes as we tune the degree of order of the resulting geometric structure. We show that when $b=1$
the system enjoys the maximum degree of freedom in the sense that cut points are chosen with uniform probability 
density and so is the delete process, since the two processes are interlocked via dimensional consistency. 
For $b>1$, at each time event a cut point is more likely to be chosen in the middle than on either end of
a given interval. That is, the frequency curve of the cut density about any given interval size is
 Gaussian in nature and as $b$ increases
the middle point is increasingly more likely to be the cut point than any other points in the interval. The
$b$ value is intimately related to the variance of the probability density of cut. As $b$ increases, the
variance of the probability density curve decreases in such a manner that in the limit $b \longrightarrow \infty$,
the variance of the curve reduces to an infinitesimally narrow line. The equivalence of the $\delta$-function and the
Gaussian model in the $b \longrightarrow \infty$ limit does indeed prove this. Changing the $b$ value
effectively means we are creating an increasingly ordered set. In this work we have derived a spectrum of 
fractal dimensions between $d_f=0.414213$ for $b=1$ and $d_f= 0.45699956$ when $b \longrightarrow \infty$. Fig. (2)
and a detailed numerical survey confirms that the fractal dimension increases monotonically with increasing order 
and reaches its constant value when the randomness is completely ceased.

In summary,  we have discussed the scaling theory of a particular class of {\it cut and delete} process, 
emphasizing on the dimensional analysis. In order to make the scaling theory more meaningful than what a simple 
scaling can provide we have invoked the concept of fractal analysis. We have identified a new set with 
a wide range of subsets produced by tuning the degree of randomness only. We have
quantified the size of the resulting set obtained in this way by fractal dimension and have
shown that the fractal dimension increases with increasing order and reaches its
maximum value when the pattern described by the set is perfectly ordered, which is
contrary to some recently found results \cite{kn.brilliantov}.  Finally, we argue on the 
basis of our findings that fractal dimension, degree of order and the extent of 
ramifications of the arising geometric patterns are intimately connected to each other.

\vspace{0.5cm}

\noindent
Acknowledgment

\vspace{0.25cm}

\noindent
M. Kamrul Hassan would like to thank  Alexander von Humboldt foundation for financial support under
the Georg Forster Fellowship.

\end{document}